\documentclass[aps,twocolumn,showpacs,floatfix,superscriptaddress]{revtex4}
\usepackage{graphicx}
\usepackage{bm}
\begin{document}
\title{Decoherence and Quantum Walks: anomalous diffusion and ballistic tails}
\author{N.V. Prokof'ev}
\affiliation{Department of Physics, University of Massachusetts,
Amherst, MA 01003, USA}
\affiliation{BEC-INFM, Dipartimento di Fisica,
Universita di Trento, Via Sommarive 14, I-38050 Povo, Italy}
\author{P.C.E. Stamp}
\affiliation{Department of Physics \& Astronomy, University of British Columbia,
Vancouver, BC, Canada V6T 1Z1}
\affiliation{Pacific Institute of Theoretical Physics, University of British Columbia,
Vancouver, BC, Canada V6T 1Z1}

\begin{abstract}
The common perception is that strong coupling to the environment
will always render the evolution of the system density matrix
quasi-classical (in fact, diffusive) in the long time limit. We
present here a counter-example, in which a particle makes quantum
transitions between the sites of a $d$-dimensional hypercubic
lattice whilst strongly coupled to a bath of two-level systems which
'record' the transitions. The long-time evolution of an initial wave
packet is found to be most unusual: the mean square displacement
$n^2$ of the particle density matrix shows long-range ballitic
behaviour, with $\langle n^2 \rangle \sim t^2$, but simultaneously a
kind of weakly-localised behaviour near the origin. This result may
have important implications for the design of quantum computing
algorithms, since it describes a class of quantum walks.
\end{abstract}
\maketitle

One can think of the trajectory of a quantum particle hopping
between 2 nodes A and B on some lattice or 'graph', as a 'quantum
walk', in which the amplitude to go from A to B is given by summing
over all possible paths (or 'walks') between them. Amusingly, such
walks can also describe the time evolution of quantum algorithms,
including the Grover search algorithm and Shor's algorithm. One can
find explicit mappings between the Hamiltonian of a quantum computer
built from spin-$1/2$ 'qubits' and gates, and that for a quantum
particle moving on some graph\cite{farhi98,kempe03}. Each graph node
represents a state in the system Hilbert space, and the system then
walks in 'information space'. This mapping is most transparent for
spatial search algorithms with the local structure of the database.
Amongst the graphs so far studied are 'decision
trees'\cite{farhi98,childs02,childs02a} and
hypercubes\cite{ambainis,childs04}; quantum walks on other graphs,
and their connection to algorithms, were recently
reviewed\cite{kempe03}.

The quantum dynamics between two sites A and B on a given graph is
often much faster (sometimes exponentially faster) than for a
classical walk on the same graph \cite{childs02a,shenvi03,kempe02}.
It has been argued that quantum walks may generate new kinds of
quantum algorithm, which have proved very hard to find.  Several
recent papers have also considered experimental implementations of
quantum walks for quantum information
processing\cite{milburn02,fuji05}; some involve walks in real space,
whereas others are purely computational (eg., a walk in the Hilbert
space of a quantum register\cite{fuji05}). Many experiments over the
years, particularly in solid-state physics, have also been
implicitly testing features of quantum walks.

As always, the main problem confronting any quantum algorithm is
environmental decoherence - the gradual entanglement of the system
with the 'environment' means that phase interference effects are
gradually lost, in measurements performed on the system alone. It is
generally assumed that the system dynamics will then show classical
diffusion at long times\cite{dec}, at least if the environment is at
or near equilibrium\cite{eqlbm}. This 'folk theorem' is supported by
results on many models\cite{weiss} (except for certain very unusual
1-dimensional systems\cite{zotos}). Recent investigations of
decoherence effects on quantum walks\cite{shapira03}-\cite{solonev}
give similar results, although in one investigation of random walks
driven by coin-tosses\cite{brun03}, non-classical behavior was
found. In these recent investigations, the decoherence mechanism was
either (i) an external noise source (ii) a coupling to a set of
tossing 'coins'; or (iii) a coupling of the coins to a heat bath. In
solid-state and atomic qubits systems, the heat bath is modelled
either by a set of oscillators (representing delocalised modes like
phonons, photons, or electrons), or by a set of '2-level systems',
or 'TLS' (representing localised modes like defects, topological
disorder, or nuclear and paramagnetic spins). Both are important in
experiment; TLS are particularly important for decoherence in
magnetic\cite{WW}, superconducting\cite{Naka}, and
conducting\cite{imry} qubit systems, and tend to dominate at low
temperature.

In this paper we consider a class of quantum walk models having a
very unusual dynamics- not only is the long-time behaviour not
classically diffusive, but a part of the single-particle reduced
density matrix always continues to show coherent dynamics. These
models are very relevant to solid-state quantum information
processing systems, since they involve a TLS bath- we couple a
quantum particle moving on a graph to a bath described by a set $\{
\mbox{\boldmath $\sigma$}_k \}$ of TLS, written as spin-1/2 Pauli
spins (with $k = 1,2,...N$). We first describe the dynamics of these
models, and then their physical interpretation.

\vspace{2mm}

{\bf Quantum Walker}: For definiteness we choose a $d$-dimensional
hypercubic graph for the walking particle (our main conclusions do
not depend on this assumption), with the 'bare' Hamiltonian
\begin{equation}
{\cal H}_o = \Delta_o \sum_{<ij>} (c_i^{\dagger} c_j +h.c.)
\label{Ho}
\end{equation}
Here $c_i^{\dagger}$ creates the particle on site $i$, and $<ij>$
restricts the dynamics to nearest neighbor hopping. The particle
moves in a Bloch band with dispersion relation $\epsilon_o ({\bf k
}) = 2 \Delta_o \sum_{\mu =1 }^{d} \cos (k_\mu a_o)$ and bandwidth
$W_o = 4d\Delta_o$. Here $a_o$ is the lattice constant, and ${\bf k}$
the d-dimensional momentum. Henceforth we measure all distances in
units of $a_o$, and label lattice sites by a lattice vector ${\bf n}$.

For this quantum walker, the solution of Schr\"{o}dinger's equation
is standard. Thus a particle initially localized at the origin, with
wave-function $\psi_{n}(t=0) = \delta_{\bf n0}$ at $t=0$, evolves
to $\psi_{n}(t) = L^{-d}\: \sum_{\bf k} e^{i[{\bf k \cdot n}
-\epsilon_o ({\bf k })]t}$ at a later time, where $L$ is the linear
system size. The probability distribution $P_{n0}^o(t) = \vert
\psi_{n}(t) \vert^2$ is then
\begin{equation}
P_{n0}^o(t) = \prod^d_{\mu = 1} J_{n_{\mu}}^2(z)\;;\; \;\;\;
z=2\Delta_o t \;, \label{Pn0}
\end{equation}
where $J_n(z)$ is the $n$-th order Bessel function. The
continuum-space limit is recovered by considering a broad Gaussian
initial wave-packet, initially centred at the origin, of form $
\psi_{ n} (t=0) \approx (1/\sqrt{\pi}R)^{d/2}\: e^{-n^2/2R^2}$
with $R\gg 1$. Then for later times
\begin{equation}
P_{ n0}^o (t) \approx \left( \frac{R^2}{\pi (R^4+z^2)}
\right)^{d/2}\: e^{-n^2R^2/(R^4+z^2)} \;. \label{PnG0}
\end{equation}
As expected, a purely quantum-mechanical evolution gives
$P_{00}^o(t) \propto 1/t^d$ and $\langle n^2 \rangle \propto t^2 $
at long times.

\vspace{2mm}

{\bf Environmental decoherence}: Coupling the quantum walker to an
environment is supposed to change the long-time evolution to
classical diffusion, characterized at long times by
$P_{00}^{(cl)}(t) \propto 1/t^{d/2}$ and $\langle n^2_{cl} \rangle
\propto t $. We certainly expect this for models in which the
particle coordinate is coupled to an Ohmic oscillator bath, but we
now examine the effect of a coupling between the particle and a TLS
bath. On its own, this bath has a Hamiltonian ${\cal H}_{TLS} =
\sum_k {\bf h}_k \cdot  \mbox{\boldmath $\sigma$}_k + \sum_{kk'}
V_{kk'}^{\alpha\beta} \sigma_k^{\alpha}\sigma_{k'}^{\beta}$, where
the $\{ {\bf h}_k \}$ are fields acting on each TLS, and the
$V_{kk'}^{\alpha\beta}$ describe interactions between them.
Typically the $V_{kk'}^{\alpha\beta}$ are very small, and lead only
to a very slow dynamics of the TLS bath, so we shall drop
them\cite{Vkk}. Various couplings of the bath to the walker are
possible, but we are interested in those which monitor transitions
of the walker, ie., those triggered when the particle hops
between nodes. We can then distinguish 2 important limiting cases:

(i) the TLS bath is acted on by only weak external fields, which we
then neglect. Now assume that each time the quantum walker hops it
can flip the $k$-th TLS $ \mbox{\boldmath $\sigma$}_k$ with
amplitude $\alpha_k$. We can write the effective Hamiltonian as
\begin{equation}
{\cal H} = \Delta_o \sum_{<ij>} \left\{ c_i^{\dagger} c_j \: \cos
\left( \sum_k \alpha_k \sigma_k^x \right) + H.c. \right\} \;,
\label{Htop}
\end{equation}
In what follows we will assume that the individual $\alpha_k$ are
small but that the number $N$ of TLS is so large that $\kappa =
\sum_k \alpha_k^2 \gg 1$, i.e. hopping events are accomplished by
simultaneous transitions in a large number of TLS. In other words,
we look at the case of {\it strong decoherence}.

(ii) The TLS bath is polarised by strong external field ${\bf
h}_k={\bf h}$. Defining the unit vector $\{ \hat{z} \}$ along the axis of
this field, and the total polarisation $M = \sum_k \sigma_k^z$ of
the TLS bath with respect to this axis, we see that in this
strong field limit, only bath transitions which conserve $M$ are
allowed. In this case one has an effective Hamiltonian
\begin{equation}
{\cal H}_M = \Delta_o \sum_{<ij>} \left\{ c_i^{\dagger} c_j \:
[P_{-M} \:e^{i \sum_k \alpha_k \sigma_k^x }\:P_M ] + H.c. \right\}
\label{Hproj}
\end{equation}
where $P_M$ projects the state of the TLS bath onto the subspace
with polarization $M$. We have dropped the large Zeeman term $\sum_k
{\bf h} \cdot  \mbox{\boldmath $\sigma$}_k$ from this Hamiltonian,
since it is now just an $M$-dependent constant.

We now proceed with the solution for the probability distribution
$P_{ n0}(t)$. We shall look in detail at the first model
(\ref{Htop}) above, and then comment on the second one. The form of
(\ref{Htop}) is a simple generalisation of a Hamiltonian ${\cal H} =
\Delta_o \left\{ \hat{\tau}_x \cos [ \sum_k \alpha_k \sigma_k^x ] +
H.c. \right\}$, which describes one limiting case of the interaction
of a single qubit $\mbox{\boldmath $\tau$}$ with a spin bath. The
density matrix of this model is given exactly as a phase average
over the propagator of the 'bare' qubit \cite{RPP,PS93}, and one can
use precisely the same technique to write the solution for
(\ref{Htop}). Thus, for the initially localised state 
$\psi_{n}(t=0) = \delta_{\bf n0}$, assuming the strong decoherence limit
described above, one finds the solution at time $t$ as
\begin{equation}
P_{n0}(t) = \int_0^{2\pi } \frac{d \varphi }{2 \pi}\prod^d_{\mu
= 1} J_{n_{\mu}}^2(z\cos \varphi  )\;,
 \label{Pn}
\end{equation}
and similarly for the initially broad wave-packet one gets
\begin{equation}
P_{ n0}(t) \approx  \int_0^{2\pi } \frac{d \varphi }{2 \pi} \;
\frac{R^d\; e^{-n^2R^2/(R^4+z^2\cos^2\varphi )} }{\big[ \pi
(R^4+z^2\cos^2\varphi)\big]^{d/2}}\:
  \;. \label{PnG}
\end{equation}
We will rederive this result using a rather different method at the
end of the paper. Given the strong coupling to the TLS environment,
one intuitively expects classical diffusive dynamics at long times.
Surprisingly, the actual evolution is radically different. Consider
first the probability at long times of finding a particle back at
the origin, $P_{00}(z \to \infty )$, in Eq.~(\ref{Pn}). The
asymptotic expansion for the Bessel function $J_0(z \cos \varphi
)\approx \sqrt{2/\pi} \cos(z \cos \varphi -\pi/4)/(z \cos \varphi )$
is not possible because $\cos \varphi \to 0$ for $\varphi \to \pm
\pi/2$. In fact, in the $t \to \infty $ limit the dominant
contribution (for $d>1$) comes from $\varphi \approx \pm \pi/2$.
Then
\begin{equation}
P_{00}(z \to \infty ) \approx \frac{1}{\pi}\int_{-\infty}^{\infty}
d\varphi J_0^{2d}(z \varphi ) = \frac{A_d}{\Delta_o t } \;,
\label{Pt0}
\end{equation}
where $A_d=(2\pi)^{-1}\:\int_{-\infty}^{\infty} dx J_0^{2d}(x)$ is a
constant (in $d=1$ there is an additional $\ln(2\Delta_o t)$
factor). This result is already rather peculiar since in $d>2$ the
decay of $P_{00}(t)$ is integrable both in the classical diffusion,
$P_{00}^{(cl)} \propto t^{-d/2}$, and in the ideal, or ballistic,
quantum propagation, $P_{00}^o \propto t^{-d}$. We get qualitatively
similar answers for the broad initial state (\ref{PnG}), where
\begin{equation}
P_{00}(z \to \infty ) \approx  \frac{A_dR^{2-d}}{ \Delta_o t } \;,
\label{Pt0G}
\end{equation}
and $A_d=2\pi^{-(1+d/2)} \int_0^{\infty} dx/(1+x^2)^{d/2} $. The
dependence on the initial wave-packet spread in the long time-limit
is another unusual feature of the solution.

From the divergence of the total time spent at the origin $\tau =
\int_0^{t\to \infty} d\tau P_0(\tau ) \propto \Delta_o^{-1}\ln
(\Delta_o t) \to \infty$ one's first suspicion is that the strong
environmental decoherence is causing some kind of quasi-localization
of the particle, analogous to weak localisation in solid-state
physics. It then comes as an astonishing paradox that a calculation
of the mean-square displacement from (\ref{Pn}) gives
\begin{equation}
<(({\bf n}(t) - {\bf n}(0))^2 >  = {1}{2}\sum_n {\bf n}^2 P_{\bf
n0}(z) = \frac{d}{2} (\Delta_o t)^2 \;, \label{n2}
\end{equation}
which is only a factor of two smaller then the coherent quantum
evolution! Thus the solution shows quasi-localsation near the
origin, coexisting with coherent ballistic dynamics at large
distances.

Having both $P_{00}(t) \propto 1/t$ and  $\sum_{\bf n} {\bf n}^2
P_{n0}(t) \propto  t^2$ at the same time is obviously
inconsistent with the simple scaling form $t^{-d}f(n^2/t^2)$. The
solution to the paradox requires a more complex shape for the
distribution function, which we show in Fig. 1 and derive here for
the Gaussian initial state. We introduce new variables $r=n/R$ and
$u=z/R^2$ to simplify the integral in (\ref{PnG}) to:
\begin{equation}
P_{r0}(u) =  \left( \frac{1}{\pi R^2} \right)^{d/2}  \int_0^{2 \pi }
\frac{d\varphi}{2\pi}\; \frac{e^{-r^2/(1+u^2\cos^2
\varphi)}}{[1+u^2\cos^2 \varphi)]^{d/2}} \;. \label{shape1}
\end{equation}
It is straightforward at this point, by considering the long-time
limit $u \gg 1$, to derive the following relations for the
intermediate
\begin{equation}
P_{n0}(t) \approx  \frac{\Gamma\left( \frac{d-1}{2}\right)
}{2\pi^{d/2+1}} \; \frac{R}{\Delta_ot\:n^{d-1}}\;,\;\;\;(R \ll n \ll
\Delta_ot/R)\;, \label{shape2}
\end{equation}
and large ($n \to \infty$) length scales
\begin{equation}
P_{n 0 }(t) \approx  \frac{1}{\pi^{(d+1)/2}n} \left(
\frac{R}{2\Delta_ot} \right)^{d-1}  e^{-n^2(R/2\Delta_ot)^2} \;.
\label{shape3}
\end{equation}
As expected from (\ref{Pt0G}), one has an increased probability of
finding a walker at the origin. Since the power-law decay
$1/n^{d-1}$ is not integrable, the normalization integral and
$\langle n^2 \rangle$ are still determined by the parameter $z$, but
the probability of being at the origin is enhanced over that at a
distance $\Delta_o t /R$  by a factor $(\Delta_ot/R)^{d-1}$.

%%%%%%%%%%%%%%%%%%%
\begin{figure}
\centerline{\includegraphics[angle=-90, scale=0.31] {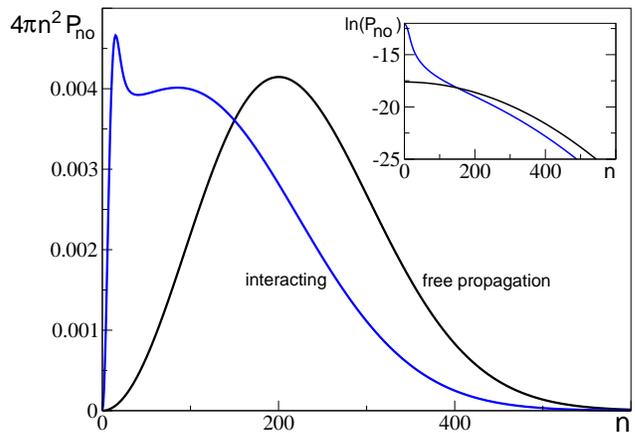}}
\vspace*{-0.4cm} \caption{(Color online). Form of $4\pi n^2 P_{
n0}(t)$ after time $t$ such that $z=2\Delta_o t \gg R^2$, calculated
from Eq.(\ref{shape1}) with $z=2000$ and $R=10$ for the
three-dimensional walker. The inset for $ln P_{ n0}$ shows the
asymptotic decay. } \label{fig1}
\end{figure}
%%%%%%%%%%%%

Not surprisingly, the thermodynamics of this system is also
peculiar. The partition function $Z=\int_0^{2\pi} (d\varphi /2\pi)
I_0^d(2\Delta_o \cos \varphi /T)$ is leading to a free energy in the
low-temperature limit $T \ll \Delta_o$ given by $F(T) \sim
-T(d/2+1/2)\ln T + const$.  

Finally, let us note that the strong-field Hamiltonian (\ref{Hproj})
gives similar behaviour. For strong decoherence we find that an
initially localized state at the origin propagates as
\begin{equation}
P_{ n0}(z) = \int_0^{\infty } dy \:e^{-y}\: P_{ n0}^0(z
J_M(2\sqrt{\kappa y})) \;.  \label{Pn'}
\end{equation}
and analysis of this shows the same long-time features as above.

\vspace{2mm}

{\bf Physical interpretation}: A path integral analysis provides
some insight here. The anomalous short-distance behaviour arises
because the effective interaction between the 2 paths of the density
matrix, generated by interactions with the spin bath environment,
has long-time memory effects in it - this is because the bath has a
degenerate energy spectrum (this is reminiscent of weak
localisation\cite{wkLoc}). But then how can we explain the
long-range ballistic tail? Usually even very weak interaction with a
bath gives classical diffusion at long ranges, because the
environment 'measures' the position of the particle as it travels
along a given path\cite{AIS90}. For this the environment does not
have to record {\it all} possible trajectories of the particle - it
only needs to track a 'coarse-grained' trajectory\cite{2-n}. The
same is true if the environment couples to the particle velocity,
from measurements of which one can also reconstruct its trajectory.

The answer to the paradox is interesting. Notice that in
(\ref{Htop}) and (\ref{Hproj}) the environmental coupling does not
distinguish different particle positions in the space of the graph
(ie., between different graph nodes), nor the direction of
transition between them; it only records that transitions between
them have occurred. This leaves room for the constructive
interference of many very large paths on the graph.

To gain more insight into the problem we rewrite the Hamiltonian
(\ref{Htop}) in the momentum representation for the walker and a
rotated basis for the TLS spins (rotating $\sigma_k^x \to
\sigma_k^z$). In this basis the Hamiltonian is diagonal; writing
$\phi = \sum_k \alpha_k \sigma_k^z$, and given some TLS spin
distribution $\{\sigma_k^z \}$ with a given $\phi$, then ${\cal H}$
acts on the eigenstates $|{\bf k}, \{\sigma_k^z \} \rangle$
according to
\begin{equation}
{\cal H} \; |{\bf k}, \{\sigma_k^z \} \rangle \;\;=\;\; \cos \varphi
\: \epsilon_o ({\bf k }) \; |{\bf k}, \{\sigma_k^z \} \rangle \;,
\label{diag}
\end{equation}
If we now start from the initially localized state for the walker 
and arbitrary $\prod_{k} | \sigma_k^z \rangle$ state for the environment
(in the original, unrotated basis) the system is an equal-weight 
superposition of all eigenstates. We immediately see that states with the same
$\varphi$ evolve coherently with a renormalized hopping amplitude
$\Delta_o \cos \varphi $, and in the strong coupling limit all
values of $\varphi$ on the $[0,2\pi]$ interval are equally
represented (thus we rederive the result given above in eqtns.
(\ref{Pn}) and (\ref{PnG})). The ballistic long-time behaviour comes
from those portions of this mixture with $\vert \cos \varphi \vert
\sim 1$. The anomalous 'sub-diffusive' long-time behavior at the graph origin, 
on the other hand, comes from a small fraction $ \sim 1/z$ of states having
very small effective $\Delta_o \cos \varphi < 1/t$, which cannot
propagate anywhere at all!

In quantum information processing systems, where the walk can occur
in different kinds of information space, no general principle forces
the environmental couplings to distinguish either the different
graph nodes, or the direction of transition between them. Thus we
see that in the design of quantum computers and certain search
algorithms, it becomes of considerable interest to look at quantum
walkers for which environmental decoherence may even be strong,
provided it is not projecting particle states onto either the
'position' or 'momentum' bases in the information space defined by
the graph on which the walk takes place. More generally, we see that
there is an interesting class of systems for which the long-time
behaviour is very far from diffusive, even in the strong decoherence
limit- instead, it combines a short-range 'sub-diffusive' behaviour
with long-range coherent dynamics.

NP thanks the BEC-INFM center at the University of Trento for
hospitality and support, and PCES thanks NSERC, PITP, and the CIAR
for support. We also thank A Hines, G Milburn, and particularly AJ
Leggett, for very illuminating discussions of these results.


\begin{thebibliography}{99}
\bibitem{eqlbm} We assume the 'quantum walker $+$ environment' to be a closed system,
                  ie., we exclude external driving fields.
\bibitem{Vkk}     Interactions between TLS in solids are usually
                  dipolar and very weak compared to other energy scales
                  (eg., in magnets, the interaction even between between
                  nearest-neighbour nuclear spins is $\sim
                  10^{-8}K$, compared to typical magnetic energy
                  scales $\sim 10-1000K$).
\bibitem{2-n}     The number of walks $N_W$ between 2 sites a distance $n$
                  apart is exponential in $n$.
                  Suppose a set of $N_M \ll n$ different devices
                  with Hilbert space dimension ${\cal D}_H$ makes
                  position measurements on the walker, recording up
                  to ${\cal D}_H^{N_M}$ possible coarse-grained paths
                  (and destroying interference between them). If
                  ${\cal D}_H$ is not exponential in $n/N_M$ (a typical
                  coarse-grained position measuring environment would 
                  have $N_M \propto n^{\beta}$, with $\beta < 1$), then 
                  ${\cal D}_H^{N_M} \ll N_W$ for sufficiently large $n$.
\end{thebibliography}
\end{document}